# Hyperspectral interference tomography of nacre

Jad Salman[1], Cayla A. Stifler[2], Alireza Shahsafi[1], Chang-Yu Sun[2], Steve Weibel[3], Michel Frising[1], Bryan E. Rubio-Perez[1], Yuzhe Xiao[1], Christopher Draves[3], Raymond A. Wambold[1], Zhaoning Yu[1,2], Daniel C. Bradley[2], Gabor Kemeny[3], Pupa U. P. A. Gilbert[2,4,5,6,7], Mikhail A. Kats[1,2,5]

[1]Department of Electrical and Computer Engineering and [2]Department of Physics, University of Wisconsin–Madison, Madison WI 53706. [3]Middleton Spectral Vision, Middleton, WI 53562. [4]Department of Chemistry, [5]Department of Materials Science and Engineering, and [6]Department of Geoscience, University of Wisconsin–Madison, Madison WI 53706. [7]Lawrence Berkeley National Laboratory, Berkeley, CA 94720.

**Abstract**

Structural characterization of biologically formed[1–7] materials is essential for understanding biological phenomena and their environment, and generating new bio-inspired engineering concepts. For example, nacre—formed by mollusks in the ocean—encodes local environmental conditions throughout its formation[8,9] and has exceptional strength due to its nanoscale brick-and-mortar structure[10,11]. This layered structure, comprising transparent aragonite tablets bonded with an ultra-thin organic polymer, also results in stunning interference colors. Existing methods of structural characterization of nacre rely on some form of cross-sectional analysis, such as scanning electron microscopy or polarization-dependent imaging contrast (PIC) mapping[8]. However, these techniques are destructive and too time- and resource-intensive to analyze large sample areas. Here we present an all-optical, rapid, and non-destructive imaging technique—*hyperspectral interference tomography* (HIT)—to spatially map the structural parameters of nacre and other disordered layered materials. We combined hyperspectral imaging with optical-interference modeling to infer the mean tablet thickness and disordering of nacre layers across entire mollusk shells at various stages of development—observing a previously unknown relationship between the growth of the mollusk and tablet thickness. Our rapid, inexpensive, and nondestructive method can be readily applied to in-field studies.

**Introduction**

Complex optical phenomena can emerge from a variety of biological or bio-inspired processes, from arrays of colors in peacocks[1] and other birds[2], butterflies[3], and opals[4], to the metal-like sheen of herring[5] and unique polarization-dependent properties of jewel beetles[6] and *Pollia* fruit[7]. Nacre, or mother-of-pearl, is a prominent biologically formed mineral structure found throughout our oceans. It lines the inside of the shells formed by many mollusks, including bivalves, cephalopods, and gastropods. It features brilliant iridescent colors (Fig. 1) and is studied and emulated in part because of its outstanding mechanical performance[10,11]. The striking, colorful appearance of nacre has been a source of scientific curiosity since the days of Brewster[12], Rayleigh[13], and Raman[14,15], and is the product of optical interference resulting from multiple interface reflections as light propagates through its stratified structure comprising stacks of transparent polygonal aragonite tablets ($CaCO_3$) interspersed with organic polymer (chitin and proteins) layers[16–18] (Fig. 1a). Nacre is one of the seven ultrastructures in mollusk shells[19]. In the nacre ultrastructure, the aragonite tablets are typically 5-10 microns in diameter and hundreds of nanometers thick (200-1100 nm across all shells, and 250-500 nm in red abalone[8]), while the organic sheets are an order of magnitude thinner[16,18,20]. In columnar nacre formed by gastropods like abalone and snails (Fig. 1), co-oriented tablets are stacked on top of one another, while in sheet nacre formed by bivalves like pearl oysters and pen shells, the tablets are staggered diagonally[8] (see Movie S1 for an animation showing how co-oriented tablets are stacked in columnar nacre). Despite the significant structural and formation-mechanism differences, the thicknesses of tablets and organic layers are similar in columnar and sheet nacre, and so is the optical behavior. The resulting palette of colors is primarily dependent on the nacre tablet thickness and the viewing angle, and the optical response that yields these colors can be understood as that of a Bragg reflector[21] with disorder in the layer thicknesses, where the optical band gaps are determined by the thicknesses of the transparent layers[5,22,23]. Thus, the spectrum of light reflected from a nacre surface encodes information about its physical structure (Fig. 1b-d).

Understanding and characterizing the structure of nacre and other biomaterials has deep and surprising implications. For example, the average thickness of the tablets comprising ancient nacre can be used as a proxy for local ocean temperatures at the time of nacre formation, enabling paleoclimatology spanning hundreds of millions of years[8,9]. The structure of nacre is also an inspiration for engineered materials thousands of times stronger than the constituent materials[17,24,25]. To that end, new techniques have been developed to probe and understand the structure of nacre, including polarization-dependent imaging contrast (PIC) mapping using x-ray absorption near-edge structure (XANES) spectroscopy combined with photoelectron emission spectromicroscopy (PEEM)[8,26]. However, both PIC mapping and more-conventional characterization techniques such as cross-sectional electron microscopy result in the destruction of the sample, and are time-consuming and costly.

Here, we present a method for rapid, non-destructive, and large-scale structural characterization of disordered and nonuniform stratified thin-film materials and apply it to the analysis of nacre. Our all-optical method employs hyperspectral imaging combined with thin-film modeling to extract nacre mean tablet thicknesses (MTT) and tablet degree of disorder ($\sigma$)—defined as the standard deviation of the thicknesses—across large areas (Fig. 1e-g). This new characterization method is designated as *hyperspectral interference tomography* (HIT). We used HIT to map the structure of mollusk shells across many stages of development and identified a previously unexplored relationship between the age of the organism and the structure of the nacre layer. We investigated two particular species of mollusk, *Haliotis rufescens* (red abalone) and *Haliotis iris* (paua, or rainbow abalone; data only in the Supplementary Information), for which the aragonite tablet thicknesses lie within a range of 250 – 500 nm[8,27], however, the method is applicable to any other transparent layered structure of animal, plant, geologic, or synthetic origin.

**Nacre as an optical material**

The lateral gaps between the abutting tablets in nacre are on the order of single nanometers[18]—much smaller than the wavelength of visible light—and therefore the aragonite/organic structure behaves largely like a thin-film assembly[22], with the color at any point along the surface of a shell attributed primarily to thin-film interference effects[21,28] (Fig. 1b-d) (see Supplementary section 1). However, nacre is far from uniform in terms of both the thicknesses of tablets comprising the stack and the surface topography, which gives nacre its unique colorful appearance.

HIT simultaneously acquires the angle-dependent reflectance spectra from many areas across the shell using a grating-based hyperspectral camera and then infers the physical Bragg-like structure from these measurements (Fig. 1e, and *Methods*). In HIT, a collimated broadband light source is used to capture predominantly specular reflected light, discarding light that is scattered or reflected by inhomogeneities on the surface. A linear polarizer is positioned in front of the camera to independently capture s- and p-polarized reflectance.

The resulting measured spectra (one spectrum per pixel, per incidence angle, per linear polarization state) can then be modeled using well-understood thin-film optical techniques, obviating the need to model the much-more-complex scattering processes.

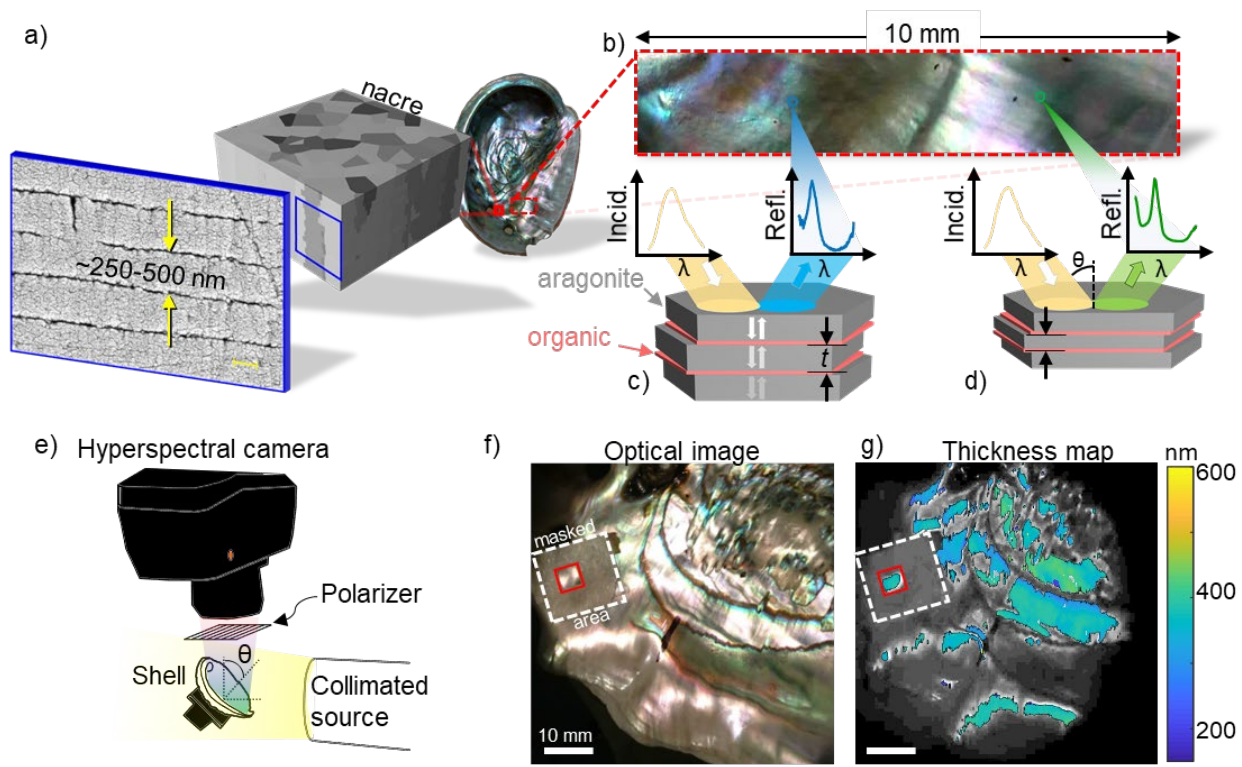

Figure 1: (a) Nacre, the colorful iridescent interior region of a mollusk shell. Here the red abalone, or *Haliotis rufescens*, shell features columnar nacre, which comprises thousands of semi-ordered layers of polygonal aragonite tablets interspersed with organic sheets. (b) A close-up photograph of the nacre surface shows a variety of colors and nonuniformities. (c, d) Given a broadband white light source illuminating nacre at a fixed angle of incidence, variations in color are observed due to the difference in average thickness of aragonite tablets comprising the nacre stack. (e) Hyperspectral interference tomography (HIT) setup: a hyperspectral camera collects predominantly specular reflectance data across a sample illuminated by a collimated source at a fixed angle of incidence from the normal to the sample ($\theta$). The reflected light is polarized using a wire-grid polarizer. (f) A color photograph of a region of nacre that was analyzed. (g) Map of the mean table thickness (MTT) obtained using HIT, overlaid on a grayscale rendering of the photograph in (f). Highlighted in red is a 5-mm-diameter region used to analyze the ontogeny of nacre in Fig. 4. This area was masked off using opaque tape, which is highlighted with the dashed white box.

**Fitting structural parameters to spectra**

At each point on the shell, the thicknesses of the tablets comprising the nacre are unknown but are expected to be normally distributed around some mean value (see Supplementary section 4). Also, the spectral reflectance measured at each pixel using HIT is a macroscopic response encompassing many stacks of nacre, because a single nacre stack area (<50 μm$^2$) is much smaller than the area represented by each measured pixel (~10,000 μm$^2$). To simulate this measurement, we performed Monte-Carlo-type calculations using the transfer-matrix method[29], which enables simulations of reflectance spectra of a thin-film assembly. We used the refractive-index values for aragonite[30] and the organic layers[31] from literature. The combined calculation yields a simulated reflectance spectrum of nacre given input variables of mean tablet thickness—MTT— and tablet degree of disorder, σ (Fig. 2a-c, see *Methods* section)[32,33].

For a given value of MTT and σ, we generated an ensemble of thin-film simulations, each comprising alternating aragonite/organic layers with thicknesses chosen stochastically from a normal distribution (Fig. 2a). We calculated the reflectance spectra for each simulation, accounting for the angular distribution of the collection optics of the hyperspectral camera (Fig. 2b). The resulting generated ensemble spectrum is the average of all simulations for a particular value of MTT and *σ*. We generated ensemble spectra for every combination of 91 values of MTT from 150 to 600 nm, and 14 values of *σ* from 5 to 70 nm (a total of 3,185,000 individual spectra were calculated for the analysis in this paper; see *Methods* for further details).

The fitting of mapped hyperspectral data to the simulated data was done pixel by pixel, as shown in Fig. 2d. Since our measurement does not capture most of the light scattered by inhomogeneities or reflected out of the acceptance cone by surface topography, the magnitude of the absolute reflectance cannot be utilized in our fitting. Similarly, since we assume a fixed number of layers in our simulations, the amplitude of the Bragg-like reflectance peaks from the simulated data cannot reliably represent the measured data. Therefore, we normalized both sets of data to their peak values. The critical spectral features are preserved, where the band peaks correspond to the MTT and the band broadening corresponds to σ[22]. A mean-squared-error (MSE) calculation was used to determine the best fit of the simulated spectrum to the measured data at each pixel (Fig. 2d center) (Supplementary section 5 for details). The associated best-fit MTT and σ are then assigned to the pixel and two maps of the shell are generated (Fig. 2d, right).

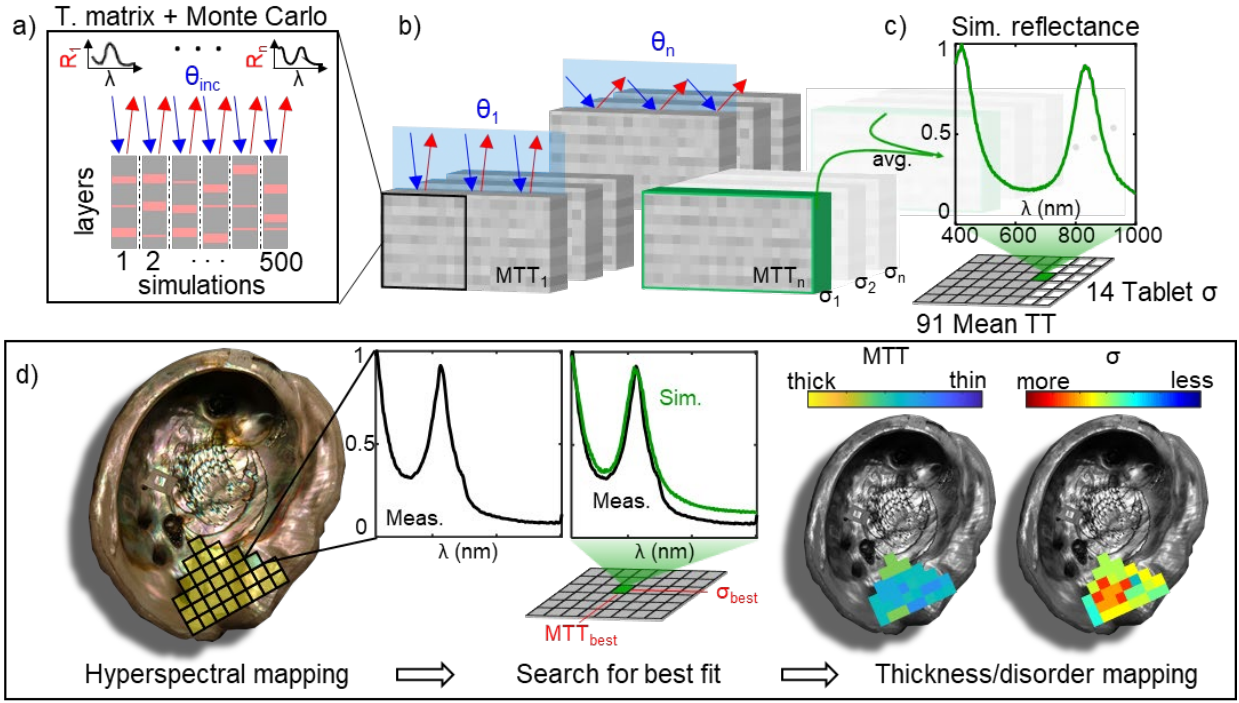

Figure 2: (a-c) Generating simulated reflectance spectra for a given mean tablet thickness (MTT) and tablet degree of disorder ($\sigma$). To generate a single simulated spectrum for a particular MTT and $\sigma$, shown in (c), a Monte Carlo method was used. (a, b) We calculated the reflectance of hundreds of randomly generated thin-film assemblies representing potential nacre layer configurations using the transfer-matrix method at various angles of incidence ($\theta_n$) present within the acceptance angle of our camera. Thousands of simulation results were averaged to produce a single reflectance spectrum, with the peak value normalized to unity. The MTT in our simulation ranged from 150 to 600 nm, with $\sigma$ ranging from 5 to 70 nm. (d) A pictorial overview of the fitting method to extract tablet thickness and $\sigma$ from measured data. Left to right: hyperspectral mapping is performed over a large region of the shell using the setup described in Fig. 1. Each pixel of the spectral map, exaggerated as yellow squares overlaid on the left image, contains a full reflectance spectrum spanning 400-1000 nm, here normalized to unity. Measured data at each pixel were then compared to every simulated spectrum generated from (c) until the lowest mean squared error was found. Each pixel was then assigned the best-fit tablet thickness ($MTT_{best}$) and disorder ($\sigma_{best}$) to generate maps of nacre tablet thickness across the entire analyzed region. Note that the maps in (d) are schematics and do not represent a physical measurement.

Although tablet-thickness disorder (encoded in $\sigma$) is presumed to be the primary cause of broadening of the peaks of the measured reflectance spectra, other physical factors are may also be captured within this metric. Any optical losses through the material or defects within the nacre would also manifest as a broadening of the Bragg peaks, but these effects are not captured in our model. Thus, we believe our extracted values of σ to be less-reliable absolute quantities compared to MTT, which is not influenced significantly by optical losses. Nevertheless, relative values of σ can be used to observe trends in tablet inhomogeneity within and between measured samples.

Due to surface curvature, topography, and scattering, only a subset of the shell can be reliably analyzed from any single measurement. This is shown in Fig. 3(a, b), where only the regions with the most specularly reflected light result in high-quality fits of the MTT (Fig. 3(b)). To identify regions with trusted fits, we defined a confidence figure of merit ($\Gamma = \frac{\text{peak measured spectral intensity}}{\text{MSE}}$). A high peak value of the raw spectral intensity indicates strong specular reflection, while a low MSE indicates that significant spectral features associated with the Bragg-like response of nacre are fit well. A map of $\Gamma$ is shown in Fig. 3(c).

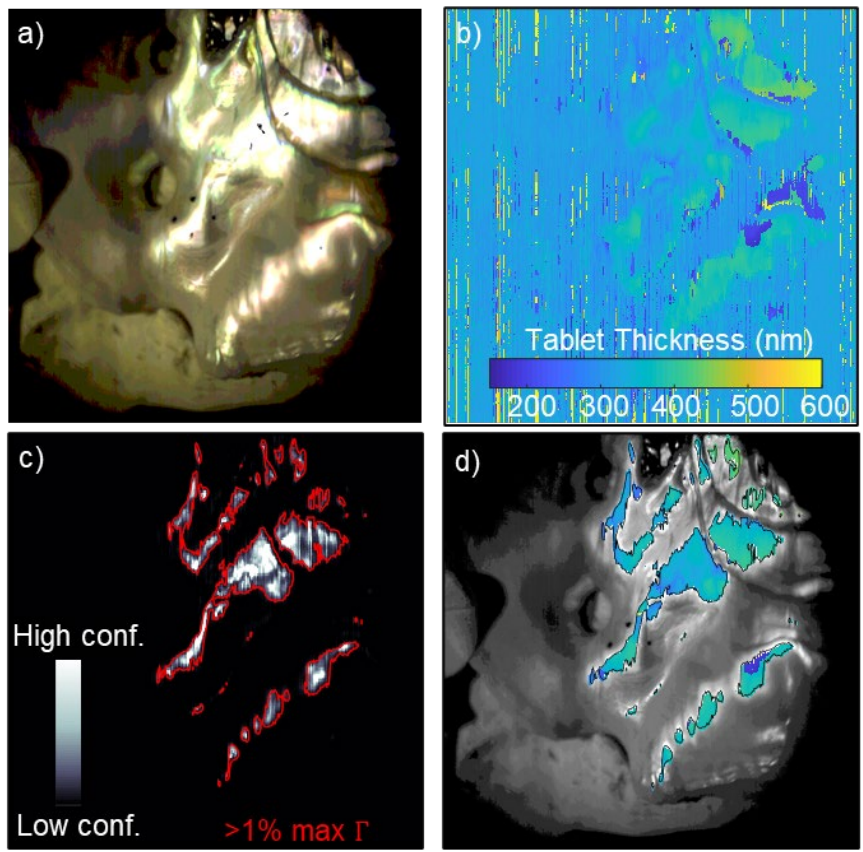

Figure 3: Interpretation of tablet thickness maps. (a) Color rendering of the hyperspectral data from a scan of a 135-mm-long red abalone shell. (b) After each pixel is fit to an MTT using the method described in Fig. 2, a full thickness map is generated. (c) Our confidence figure of merit, $\Gamma$, is used to identify regions of reliable data. To scrub unreliable data, we set a minimum threshold of $\Gamma$. (d) An overlay of our high-confidence MTT data and a greyscale image of the measured area.

σ to be less-reliable absolute quantities compared to MTT, which is not influenced significantly by optical losses. Nevertheless, relative values of σ can be used to observe trends in tablet inhomogeneity within and between measured samples.

Due to surface curvature, topography, and scattering, only a subset of the shell can be reliably analyzed from any single measurement. This is shown in Fig. 3(a, b), where only the regions with the most specularly reflected light result in high-quality fits of the MTT (Fig. 3(b)). To identify regions with trusted fits, we defined a confidence figure of merit ($\Gamma = \frac{\text{peak measured spectral intensity}}{\text{MSE}}$). A high peak value of the raw spectral intensity indicates strong specular reflection, while a low MSE indicates that significant spectral features associated with the Bragg-like response of nacre are fit well. A map of $\Gamma$ is shown in Fig. 3(c).

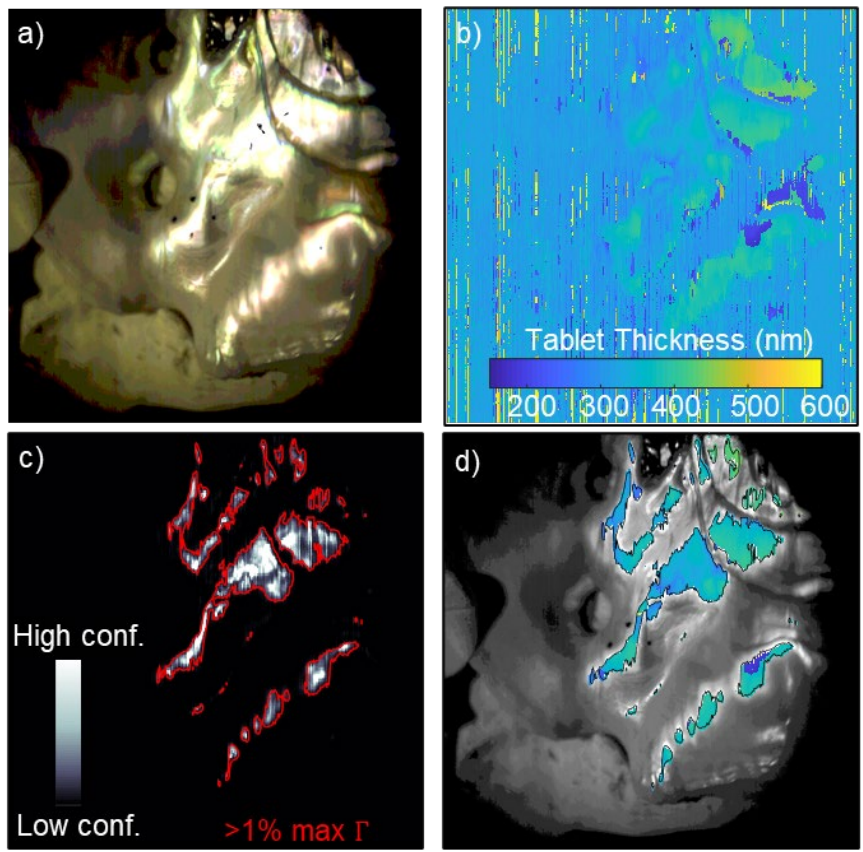

Figure 3: Interpretation of tablet thickness maps. (a) Color rendering of the hyperspectral data from a scan of a 135-mm-long red abalone shell. (b) After each pixel is fit to an MTT using the method described in Fig. 2, a full thickness map is generated. (c) Our confidence figure of merit, $\Gamma$, is used to identify regions of reliable data. To scrub unreliable data, we set a minimum threshold of $\Gamma$. (d) An overlay of our high-confidence MTT data and a greyscale image of the measured area.

Setting the Γ threshold to 1% of the maximum Γ was enough to eliminate the most unreliable data, *e.g.*, shadows, non-nacre surfaces, and non-specular regions (Fig. 3d). The same method can be applied to the nacre disorder ($\sigma$) maps (see Supplementary section 6). The expected systematic error of HIT mapping was found to be at most ~25 nm in thickness at a given angle of incidence (see Supplementary section 2). We validated the accuracy of HIT via scanning electron microscopy (SEM) measurements made on nacre cross-sections at several locations on one of the shells. The SEM and hyperspectral measurements agreed to within ~15 nm for MTT and less than 5 nm for σ (see Supplementary section 7). By simply readjusting the sample orientation and repeating the measurements, reliable data can be captured from the entire surface.

**Using HIT to investigate nacre formation**

HIT enables rapid, non-destructive characterization of spatially resolved structural parameters of nacre or any other layered material across large areas. We used HIT to investigate whether there is a relationship between mollusk developmental stage and the structural parameters of the nacre formed by the animal.

We purchased eleven pairs of live red abalone at different developmental stages—from 1 to 6 years of age—corresponding to shell lengths between 20 and 140 mm (see Supplementary section 9). The samples were all grown in the same environment by the Monterey Abalone Company (Monterey, CA, USA), in cages suspended under the Monterey Wharf and thus exposed to natural ocean water composition and temperatures. All live animals were collected on August 6, 2019. Therefore, the surfaces of the nacre across the samples were formed at the same time and at the same temperature.

For each of the 22 red abalone samples, we defined two square regions (5 mm × 5 mm): site 1 near the outer-most complete gill hole and site 2 near the fourth gill hole away from the site 1 (Fig. 4a). As the animal grows older—therefore larger—nacre is continuously deposited across the entire surface. Thus, for every sample, site 1 contains nacre formed when the mollusk was at its oldest prior to harvesting, while site 2 contains nacre formed both when the mollusk was younger and older prior to harvesting[27]. Several spectral maps were taken for each site. The MTT values for all pixels in a site map were averaged to yield a single value, *e.g.*, $\overline{\text{MTT}}$ = 400 nm in Fig. 4b (see *Methods*).

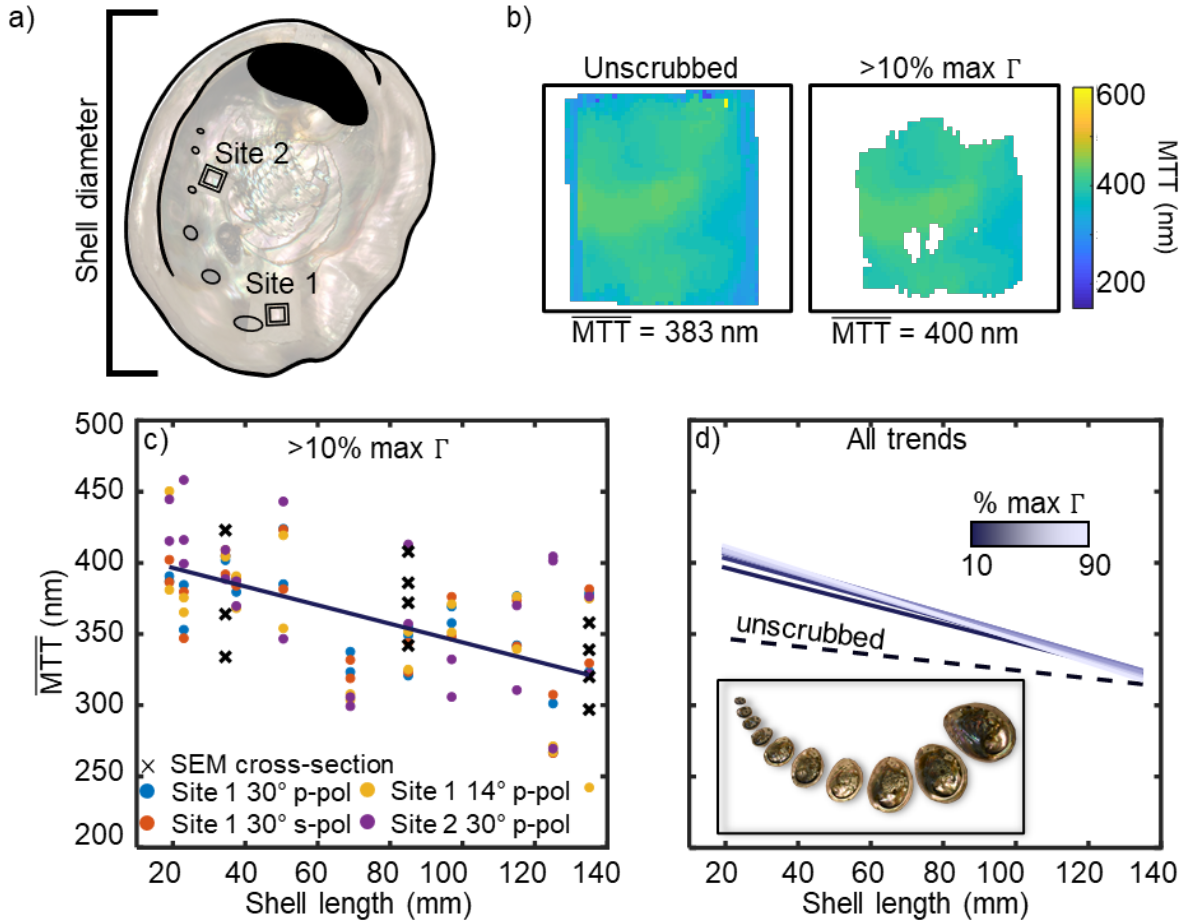

Figure 4: Ontogenetic dependence of red abalone nacre. (a) Two square measurement sites were defined on each of the 22 shells (11 pairs), each 5 mm × 5 mm. Site 1 was near the outer-most complete gill hole in each shell and contains nacre predominately formed when the mollusk is oldest. Site 2 was near the fourth gill hole inward from site 1 and contains nacre formed both when the mollusk was younger and older. (b) Mean tablet thickness (MTT) maps representing a single site on one of the shells. This region is a subset of the full 512 × 512 pixel map. (c) Plots of all site averages versus shell length, with the Γ threshold set at 10%. Each pair of shells with the same shell length has 8 thickness measurements in total, with different polarizations and incidence angles, as described in the legend. The dark line is a linear fit to all data points. (d) Plots of linear-fit trends obtained selecting different scrubbing thresholds, with the dashed line representing unscrubbed data, and the lightest purple representing scrubbed data with the Γ threshold set to 90% of the maximum Γ for each map.

The data in Fig. 4c shows an inverse relationship between $\overline{MTT}$ from all sites and shell length, and therefore age. This trend persists for all levels of data scrubbing (Fig. 4d), indicating that younger abalone grow thicker nacre tablets compared to older abalone (see Supplementary section 9), an observation that has not been previously reported in this or any other nacre-forming mollusk genus or species. In fact, because it appears to the naked eye to be mostly similar in shells of different sizes and ages, nacre has always been assumed to have the same ultrastructure, and specifically MTT[8,19,34]. A similar trend was also noted in the tablet disorder $\sigma$, with younger abalone producing more disordered nacre compared to more mature abalone (see Supplementary section 8).

The generality of the relationship between MTT and age to other nacre forming mollusks, and to other layered biominerals, including Bouligand structures in bone, shrimp clubs, fish scales, and insect cuticles[35–38], remains to be tested. Using HIT, such experimental analysis is straightforward and relatively inexpensive, thus it can be done on a vast scale by many research groups.

**Conclusion**

In conclusion, we demonstrated an all-optical, rapid, and non-destructive characterization technique—denoted as *hyperspectral interference tomography* (HIT)—for extracting structural information from disordered and nonuniform layered materials such as nacre and other biominerals. The accuracy of HIT in determining the thickness and disorder of aragonite tablets comprising nacre formed by red abalone (*Haliotis rufescens*) and rainbow abalone (*Haliotis iris*) was confirmed by electron-microscope cross-sections.

By performing large-area HIT analysis of red abalone shells at different stages of development, we observed a previously unknown dependence of the structure of nacre on the development stage of the animal. The HIT technique can be easily exported to other labs, or deployed to the field for the study of large-area, biologically formed, layered optical materials, even if they are delicate, contoured, and nonuniform in topography.

**Acknowledgements**

MAK acknowledges support from the Air Force Office of Scientific Research, award FA9550-18-1-0146. PG acknowledges 80% support from the U.S. Department of Energy, Office of Science, Office of Basic Energy Sciences, Chemical Sciences, Geosciences, and Biosciences Division, under award DE-FG02-07ER15899, and 20% support from NSF grant DMR-1603192.

**Methods**

**HIT measurement setup**

We used a hyperspectral camera (Specim IQ, Specim Ltd., Oulu, Finland), rotation stage, and a collimated broadband halogen light source (40 W, reflector-type) with a beam divergence ~2.4° to illuminate and measure reflectance spectra across large areas of nacre (Fig. 1e). The spectra were all normalized to reflectance from an aluminum mirror under the same illumination conditions. The spatial resolution of our setup is 512 x 512 pixels, with each pixel containing a full reflectance spectrum from 400 to 1000 nm with 204 spectral bands. A linear polarizer was placed in front of the camera lens to enable s- and p-polarized reflectance measurements. Due to the positioning of the hyperspectral camera, here the spatial resolution of the HIT image is ~100 μm. The approximate acceptance angle of light captured from each point on the surface of the nacre was ±9°. The use of collimated incident light and a relatively small acceptance angle was deliberate to capture predominantly specular reflected light, and discard light that is scattered or reflected by inhomogeneities on the surface.

**Transfer-matrix method combined with Monte Carlo simulations**

A transfer-matrix-method calculation requires several inputs to calculate a spectrum: the refractive indices and thicknesses for every layer and the angle of incidence, polarization, and wavelength of the incident light. The optical properties of nacre's constituent components, aragonite and the organic sheets, are approximately known. Aragonite is a biaxial birefringent material with refractive index of 1.681, 1.686, 1.530 along the *a*-, *b*-, and *c*-axes of the crystal at λ = 589 nm, respectively[30]. To our knowledge, the dispersion of aragonite has not been fully characterized in literature; however, we assumed it to be dispersionless (and lossless) across the visible and near-infrared wavelengths, as is the case for calcite[39]. Since the crystal *a*- and *b*-axis orientations of the nacre tablets can vary by ±90° from one layer to the next, and the *c*-axis orientation can vary by ±30° from the normal to the nacre layers[8,40], we selected a refractive-index value of 1.63, which is the arithmetic mean of the three values. The refractive index of the organic sheets ranges between approximately 1.4 and 1.73 across different species[41]. We selected a constant refractive index of 1.43[31], noting that the results are minimally impacted for different refractive-index values within the known range for the organic sheets (see Supplementary section 2). The fitted results for s- and p-polarized measurements shown in Fig. 4c are very similar, affirming that although aragonite is biaxially birefringent[30], the wide variation in crystal orientations throughout the nacre averages all polarization-dependent effects and therefore can be reasonably treated as a single isotropic refractive index.

Although nacre can be thousands of layers thick[9,20], we are likely measuring the first few hundred layers (see Supplementary section 3), likely due to internal scattering which prevents deeper penetration of light. Thus, our model assumes a thin-film assembly of 200 layers. Furthermore, the thicknesses of the organic sheets were not used as fitting variables and were fixed at 25 nm[16]. We set reasonable bounds for the expected MTT between 150 and 600 nm, as well as σ between 5 and 70 nm. Monte-Carlo-type calculations were performed to determine the expected ensemble spectral response of nacre from each pixel in a HIT map. For a given value of MTT and σ, 500 thin-film simulations were generated, each comprising 200 alternating layers with assigned thicknesses chosen stochastically from a normal distribution centered at the mean with a standard deviation equivalent to the tablet degree of disorder, σ (Fig. 2a). We calculated the reflectance spectra of all 500 simulations, accounting for the angular distribution of the collection optics by averaging calculated spectra over five angles spanning the acceptance angle of the camera at a single angle of incidence, *e.g.*, 30° ± 9° (Fig. 2b). The single ensemble spectrum generated is the average of all 2,500 simulations (Fig. 2c). In total, 91 × 14 ensemble spectra were calculated spanning the thickness and disorder bounds that we defined (a total of 3,185,000 spectra).

**Red abalone ontogeny measurements**

Twenty-two living red abalone at different developmental stages—from 1 to 6 years of age—corresponding to shell lengths between 20 and 140 mm were measured. The samples were all grown in the same environment by the Monterey Abalone Company (Monterey, CA, USA), in cages suspended under the Monterey Wharf and thus exposed to natural ocean water composition and temperatures. All live animals were collected on August 6, 2019. Site 1, located near the outer-most value (Fig. 4a), for all shells was measured with the following conditions: p- and s-polarized reflectance at a 30° incidence angle and p-polarized reflectance at 14° incidence. Site 2, located near the fourth valve from site 1 (Fig. 4a), for all shells was measured with p-polarized reflectance at a 30° incidence angle. The $\overline{MTT}$ was calculated for each shell as the Γ threshold was increased from 0% to 90% of the maximum. In Fig. 4c, we plotted the $\overline{MTT}$ for all shells versus the shell length with the minimum Γ threshold set to 10% of the maximum.

**SEM cross-sectional validation of nacre formation**

We cut the shells in the measured locations labeled Site 1 and Site 2 (Fig. 4a), embedded and ground them to expose the shell cross-sections at both sites (see Supplementary section 7), polished and coated them, and finally used scanning electron microscopy (SEM) to measure the tablet thicknesses directly.

The SEM measurements are reported in Fig. 4c, and in the Supplementary Information. The SEM data confirm that MTT decrease with shell size, making this a robust observation and validating our HIT results. We note that SEM only provides a highly localized measurement of the nacre structure, is invasive, destructive, and time-consuming because of the sample preparation required. Thus, a limited number of samples (135-, 85-, and 34.5-mm lengths) were prepared and analyzed with SEM (11 datapoints in Fig. 4c). All SEM data points are in excellent agreement with the trend observed with the larger-area, non-invasive, and non-destructive HIT technique.

**References**

1. Newton, I. *Opticks* (1704).

Supplementary Information:

# Hyperspectral interference tomography of nacre

Jad Salman[1], Cayla A. Stifler[2], Alireza Shahsafi[1], Chang-Yu Sun[2], Steve Weibel[3], Michel Frising[1], Bryan Rubio-Perez[1], Yuzhe Xiao[1], Christopher Draves[3], Raymond Wambold[1], Zhaoning Yu[1,2], Daniel C. Bradley[2], Gabor Kemeny[3], Pupa U. P. A. Gilbert[2,4,5,6,7], Mikhail A. Kats[1,2,5]

[1]Department of Electrical and Computer Engineering and [2]Department of Physics, University of Wisconsin–Madison, Madison WI 53706. [3]Middleton Spectral Vision, Middleton, WI 53562. [4]Department of Chemistry, [5]Department of Materials Science and Engineering, and [6]Department of Geoscience, University of Wisconsin–Madison, Madison WI 53706. [7]Lawrence Berkeley National Laboratory, Berkeley, CA 94720.

## 1. Structural color dependence on angle

Optical interference effects in thin-film assemblies are dependent on the thicknesses and refractive indices of the films and the angle of the incident light with respect to the structure. Figure S1 shows the structural color of nacre produced by two different mollusk species: *Haliotis rufescens* and *Haliotis iris*. The iridescent colors are clearly dependent on the viewing angle, as seen when the angle of incidence is changed between the illuminating source (sunlight) and the camera. As the samples are tilted from near-normal to oblique angles (Fig. S1 a-c, d-f) the range of reflected colors shift from predominately red, orange, and yellow (longer wavelengths) to greens, blues, and violets (shorter wavelengths).

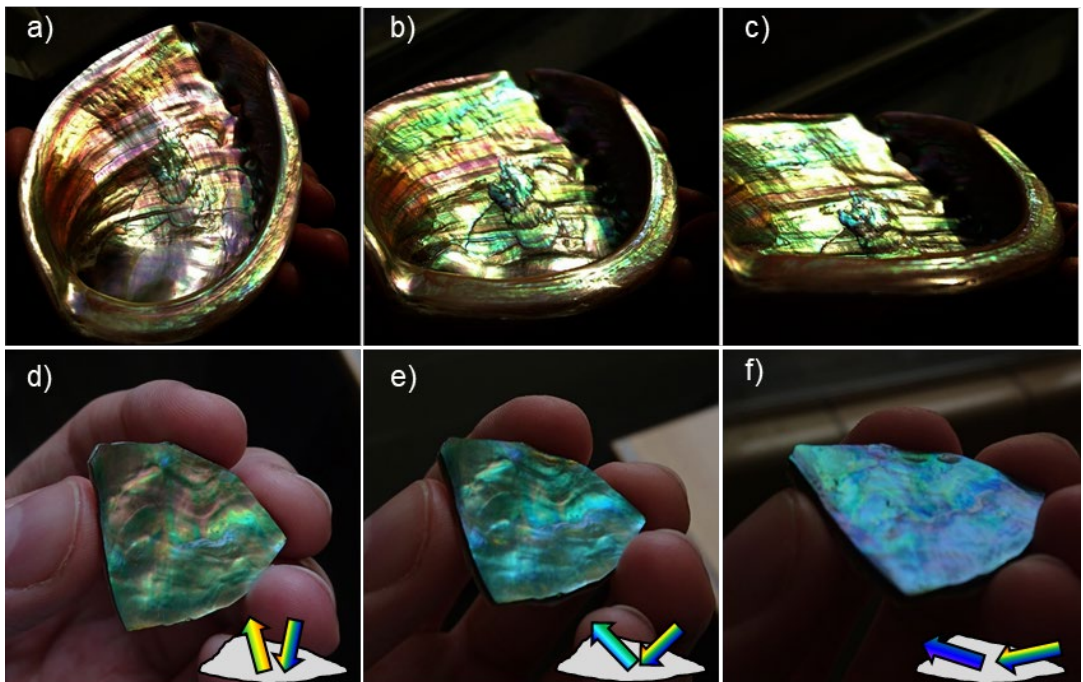

**Figure S1:** Structural-color dependence on the incidence and reflection angles. Nacre samples from *Haliotis rufescens* (a-c) and *Haliotis iris* (d-f) were imaged at various incidence angles (left to right) while illuminated with sunlight.

Interference enables the high reflectivity in Bragg reflectors. In an ideal Bragg reflector, alternating lossless dielectric materials are stacked with thicknesses corresponding to $\frac{\lambda_o}{4n_{1,2}}$ (when designed for normal incidence), where $\lambda_o$ is the central wavelength for constructive interference in reflection and $n_{1,2}$ corresponds to the refractive index of the high- or low-index layers, respectively. In Fig. S2a, the simulated reflectance spectrum for a 50-layer Bragg reflector is shown. The thin-film assembly was designed to have peak reflectance at $\lambda_o$ = 600 nm at normal incidence and used alternating high- and low-index materials equivalent to nacre's aragonite crystals ($n_1$ = 1.63) and organic polymer ($n_2$ = 1.43). Thus, the thickness of the high- and low-index layers were 92 and 105 nm, respectively. As the angle of incidence transitions from normal to oblique, the central wavelength shifts towards shorter wavelengths.

The stratified structure of nacre is essentially a Bragg reflector with disorder in the layer thicknesses. Figure S2b shows the reflectance spectrum for the same Bragg reflector as in S2a but with a 20 nm standard deviation in the thicknesses from a mean set to the ideal thicknesses for the high- and low-index layers. The reflectance spectra shown are the average of 30 simulations with randomly distributed disorder in the layer thicknesses. The angle dependence of the Bragg peak remains, though the amplitude is reduced and the peak is broadened compared to the perfect Bragg reflector.

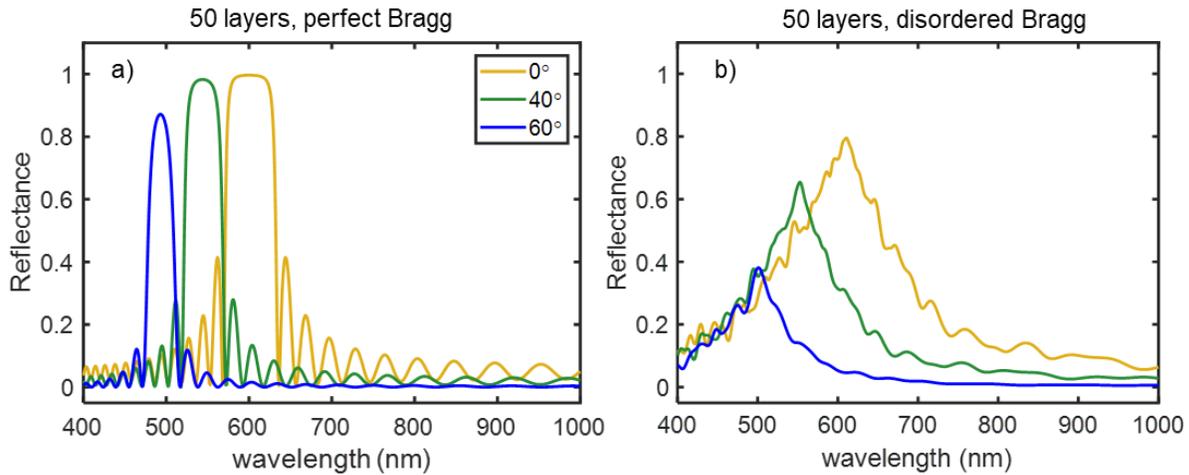

**Figure S2:** Bragg reflector without (a) and with (b) disorder. a) Transfer-matrix method calculation of the reflectance from a 50-layer perfect Bragg reflector designed with central wavelength $\lambda_o$ = 600 nm for normal incidence. The refractive indices, $n_1$ and $n_2$, are 1.63 and 1.43, respectively. b) The calculated reflectance spectrum of a 50-layer disordered Bragg reflector for which the standard deviation of the thicknesses is 20% of the perfect thickness from the ideal condition. These spectra are the average of 30 randomly generated disordered stacks.

## 2. Impact of model assumption on fitting

We assumed the constituents of nacre to be dispersionless and lossless throughout the visible and near infrared, and consistent with values reported in literature. The aragonite tablets were assigned refractive index $n_{\text{aragonite}}$ = 1.63, and the organic polymer was assigned refractive index $n_{\text{organic}}$ = 1.43. The real constituent materials likely have some dispersion across the wavelength range of interest, and a range of literature values for chitin proteins refractive indices have been reported. To determine the potential impact to our modeling of nacre, we calculated the individual impact on the final mean tablet thickness (MTT) fittings given reasonable variations in the refractive indices of the aragonite and organic layers, as well as systematic variations in the angle of incidence.

Table S1 details the impact to MTT fits due to deviations in the modeling assumptions. First, the baseline calculation sets $n_{aragonite}$ = 1.63, $n_{organic}$ = 1.43, and the angle of incidence (AOI) to 30°, with the remaining model assumptions the same as outlined in the main text. A set of reflectance spectra, $R$, are generated for all MTT (150 – 600 nm) and tablet degree of disorder, $\sigma$ (5 – 75 nm). Then, for each variation in the refractive index or AOI parameters, a new set of reflectance spectra, $R'$, is calculated within the same MTT and $\sigma$ parameter space. The impact from each individual parameter variation is determined by performing a fit (see Fig. 2 in the main text) between $R$ and $R'$ and extracting an expected MTT. The difference between baseline and the expected MTT fits is a measure of the impact of a given parameter variation on the final MTT. The largest impact on MTT was when setting $n_{aragonite}$ = 1.53, the refractive index of aragonite along the *c*-axis, with an approximately 30 nm change to the MTT. Using $R'$ values to fit the MTT using measured nacre spectra shows a similar impact on the MTT.

**Table S1:** Impact on the calculated mean tablet thickness (MTT) from model variations. The impact is quantified as the magnitude in the average of the differences in MTT.

|  | $n_{aragonite}$ = 1.53 | $n_{aragonite}$ = 1.68 | $n_{organic}$ = 1.4 | $n_{organic}$ = 1.73 | AOI = 35° | AOI = 25° |
|---|---|---|---|---|---|---|
| Simulated [a] \|ΔMTT\| (nm) | 27 | 11 | 6 | 2 | 9 | 3 |
| Measured [b] \|ΔMTT\| (nm) | 25 | 11 | 10 | 1 | 6 | 5 |

[a] The difference in MTT between the fitting of $R$ to $R'$ datasets
[b] The difference in MTT between the fitting of measured nacre reflectance to $R$ and $R'$

### 3. Optical modeling using SEM-measured layer thicknesses to determine penetration depth

Light interacting with nacre can be scattered and attenuated by inhomogeneities and losses in the material. As described in the main text, hyperspectral interference tomography (HIT) restricts the measurement to specular reflections. However, the maximum depth at which the specular light penetrates, reflects, and is detected from nacre needs to be known to build a reasonable optical model for fitting structural parameters. To determine the penetration depth of our measurement technique, we measured the reflectance spectra from a piece of nacre formed by *Haliotis rufescens*, then performed cross-sectional scanning electron microscopy (SEM) measurements of the full thickness of the nacre stack. The nacre stack was modeled about a hundred layers at a time using SEM thickness measurements for each layer—starting from the surface and ending at the prismatic layer. The modeled reflectance spectra were compared to reflectance measurement at the cross-section location to determine the penetration depth. In all, more than 3300 layers were modeled from the cross section. Figure S3 shows the normalized calculated reflectance spectra of the nacre as the number of layers is increased through the depth of the stack. The curve in black is a measured normalized reflectance spectrum from a site adjacent to the cross sectioned area. Figure S3b shows that the simulated reflectance spectra with the best fit to the measured reflectance occurs within the top 400 layers of nacre. A similar test was performed on a sectioned piece of *Haliotis iris* with the best fit occurring for the top 148 layers of nacre shown in figure S3c,d. Thus, for HIT, we assumed a uniform penetration depth of 200 layers in our optical modeling.

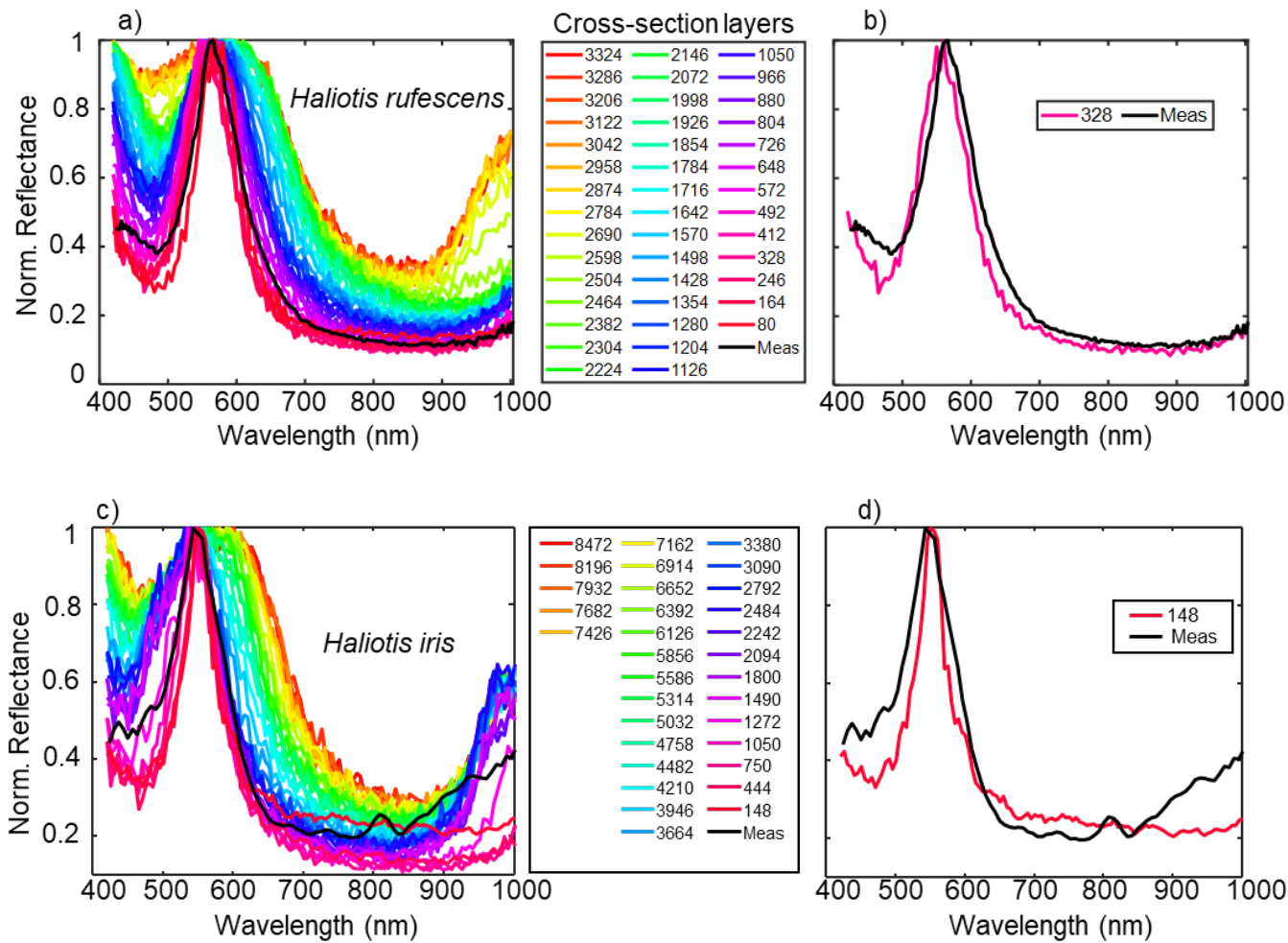

**Figure S3:** Simulation of cross-sectioned nacre thin-film assembly of Nacre. a) Transfer-matrix-method calculation of reflectance of a nacre thin-film assembly as layers are sequentially added to the model based on true cross-sectioned measurements of the tablet thicknesses formed by *Haliotis rufescens*. The first red line includes the top 80 layers of nacre, and so on until the entire nacre stack is simulated which includes over 3300 layers. In black, a normalized measurement of reflectance taken immediately adjacent to the cross-section cut. b) The simulated best-fit reflectance occurs at about 330 layers in the stack. c) Simulated reflectance based on a sectioned piece of *Haliotis iris* with the normalized measurement of reflectance at the sectioned location overlaid in black. d) The corresponding best fit for this sample occurs at about 150 layers.

**4. SEM measurements of the distribution of nacre tablet thicknesses**

The optical model used in HIT assigns the aragonite tablet thickness values randomly from a Gaussian distribution centered at a given mean value and standard deviation (given by the degree of disorder). To confirm the distribution of tablet thicknesses is indeed Gaussian, we extracted the individual tablet thicknesses via cross-sectional SEM measurements on a piece of nacre formed by *Haliotis rufescens*. Figure S4 shows the entire cross-sectioned region and highlights the area where individual layer thicknesses were measured. Image processing of the SEM data measured edge-to-edge spacing between individual tablets.

A histogram plot shows that the distribution of the tablet thicknesses at this location on the shell is approximately Gaussian, with a mean centered at 420 nm and a standard deviation of 50 nm.

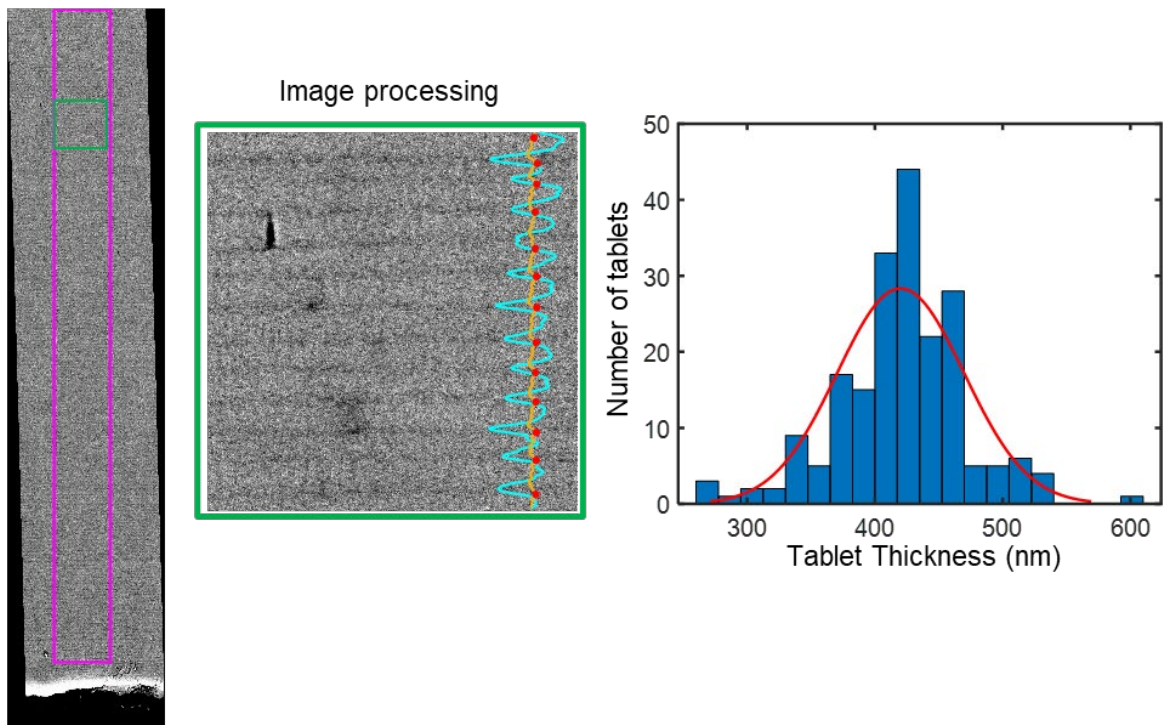

**Figure S4: Nacre tablet distribution validation.** (left) A composite cross-sectional SEM image of a nacre stack from a sample of *Haliotis rufescens*. The pink rectangle highlights the region where individual nacre tablets are counted, and thicknesses measured. (center) The inset shows a close-up region of nacre tablets defined by the green box. Image processing of the pixel intensities allow for individual tablets to be resolved by calculating edge-to-edge differences (shown as intensity profiles (cyan line), intensity derivative (orange line), and points of maximum slope (red dots)). (right) A histogram of the measured tablets in the cross-section. The tablets have a Gaussian distribution of thicknesses centered around a mean of 420 nm with a standard deviation of 50 nm

### 5. Fitting of thickness from spectral data

To a first order, nacre can be modeled as a disordered Bragg reflector. As described in the main text, we combined transfer-matrix-method calculations with the Monte Carlo method to simulate 91×14 ensemble spectra given inputs of aragonite mean tablet thickness, MTT (150 – 600 nm), and tablet degree of disorder, σ (5 – 70 nm) (see Fig. 2 in main text). Fitting was done pixel by pixel to determine the best fit between the simulated and measure spectra. The fitting procedure entails several steps and was performed computationally using MATLAB.

First, all measured and simulated spectra were normalized to their peak reflectance values, which was set to 1 (Fig. S5a, black line). The normalization eliminated discrepancies in absolute reflectance magnitudes. The most critical spectral features are preserved and correlate strongly with the MTT. The band broadening, which is an indicator of nacre tablet disorder, is also preserved. After normalization, the spectra are smoothed using a moving-average function with a 10-point average to eliminate high-frequency fluctuations in the spectra from measurement noise (Fig. S5a, gray line). Then, numerical differentiation of the smoothed spectra is calculated with respect to the wavelength. The derivative of the measured reflectance for each pixel in our hyperspectral maps is compared to the derivate of all simulated reflectances (91 × 14 simulated ensemble spectra) using a mean-squared error (MSE) calculation. The simulated spectrum that has the lowest MSE value is considered the best-fit and its corresponding MTT and σ are recorded (Fig. S5b).

A second refinement fit is done using the initial best-fit parameters as a starting point. The MSE is calculated between original measured reflectance spectra and all simulated spectra within 20 nm of the initial best-fit MTT and σ. The fit parameters that produce a spectrum with the lowest MSE are chosen as the final best fit for the specific pixel (Figure S5c).

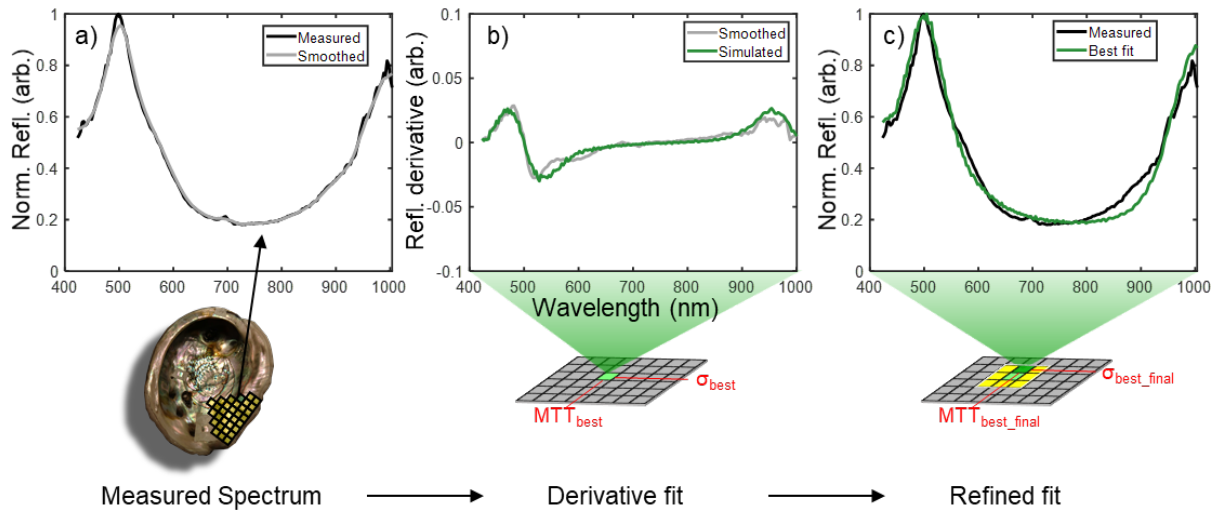

**Figure S5: Refined MTT and σ fitting method.** a) A measured reflectance spectrum from a single point along the nacre shell is normalized to the peak reflectance (black line). The spectrum is then smoothed using a 10-point moving average to reduce noise (grey line). b) The derivative of the smooth spectrum is compared to the derivatives of all 91 × 14 ensemble simulated spectra until the lowest mean-squared error is found between the two spectra. An initial MTT$_{best}$ and σ$_{best}$ is found. c) A second fitting is done within a subset of the simulation parameter space (white box with yellow squares). The space is defined as all spectra which are within 20 nm of the initial MTT$_{best}$ and σ$_{best}$ determined in (b). The final fitting is a direct comparison of normalized measured and simulated reflectance spectra. The final refined fit is selected and assigned to the point on the mapped shell.

The purpose of the double fitting method—with the first fitting based on the derivative of the reflectance and the second fitting based on the direct spectra—is to improve the reliability of fitting to points on the nacre with noisy, low-intensity spectra. As shown in Fig. S6a-c, fitting directly to spectra yielded reliable fits to measurements with clean and high-intensity spectra, *i.e.*, regions with strong specular reflectance. However, direct fitting often failed to reasonably fit to measured spectra that clearly have Bragg reflectance bands but are noisy due to a low-intensity specular component. Therefore, by first taking the derivative of the smoothed spectra, noise was reduced and baseline offsets in the spectra were removed, leaving primarily the spectral features of interest—the high-reflectance bands—for fitting. The fitting of noisy, weak reflectance spectra was significantly improved with the introduction of the derivative fit.

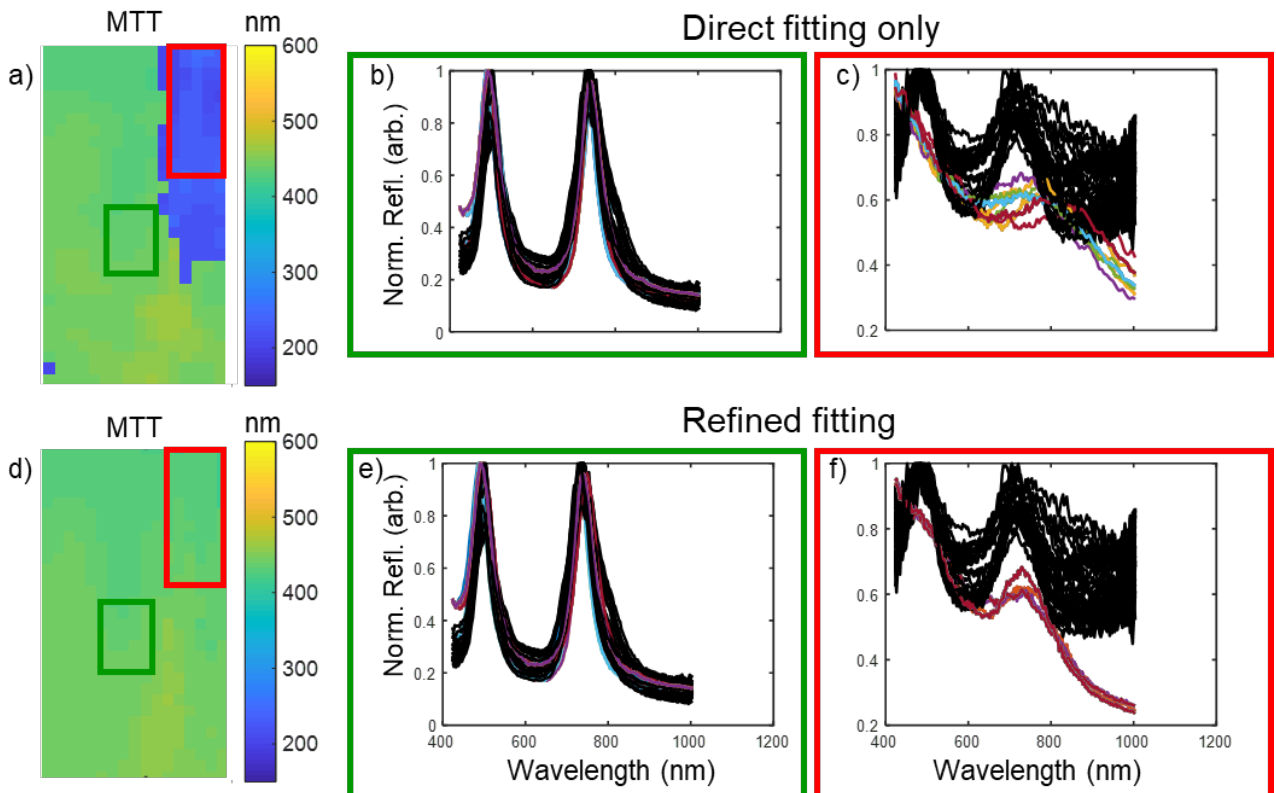

**Figure S6: Comparison of single versus refined fitting.** a) A mean tablet thickness (MTT) map based on a direct fitting of measured (black lines) and simulated (colored lines) reflectance spectra. In the map, a stark shift in the MTT is noted between the green and red highlighted regions. b) The best fits from the green highlighted region have good agreement between measured and simulated spectra. The measured spectra have clearly defined features with low noise—indicative of strongly specular reflectance. However, the red highlighted region has poor agreement with high variability in the fits due to increased noise from the weakly specular measurements. d) By fitting to the derivative of the spectra first, followed by a refined direct fit of spectra, the artificial discontinuities (dark blue regions versus light green regions) in the MTT maps are eliminated. e) High-quality fits are maintained for strongly specular data, while for noisy spectra (f), better fitting is achieved.

**6. Details of scrubbed fitting maps for thickness and disorder**

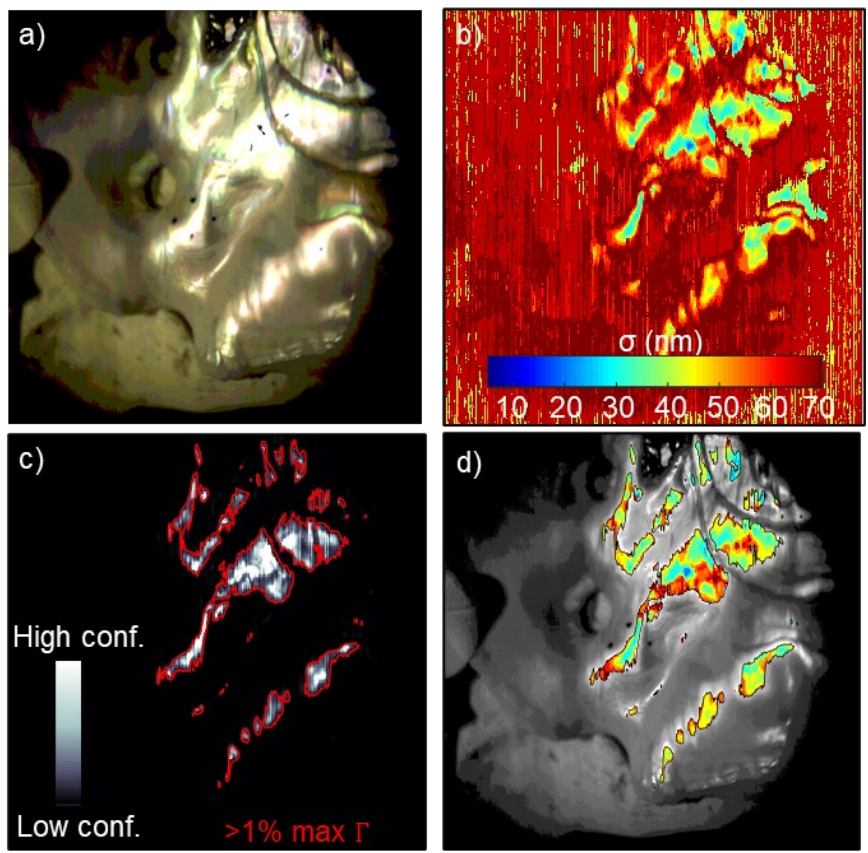

**Figure S7: Scrubbed degree of disorder (σ) maps.** a) A color rendered image of a section of a 135-mm-long red abalone shell. After fitting, the best-fit MTT (Figure 3) and σ maps (b) are extracted. c) A map of the figure-of-merit confidence, Γ, is shown. (d) An overlay of our scrubbed σ map and a greyscale photo of the measured area.

## 7. SEM validation cross-sections

Sections of the largest abalone shell were cut with a TechCut 4™ precision low-speed saw (Allied High Tech Products, Rancho Dominguez, CA, USA), being careful to preserve the two analyzed regions described in Fig. 4 in the main text. A notch was made in the shell to identify specific regions on the shell where direct comparisons could be made. After the desired cross section was cut, that section of the shell was analyzed with the hyperspectral camera and the mean tablet thickness was calculated via the methods described above. The shell was then embedded cut-side down in Solarez ® UV resin (Wahoo International, Vista, CA, USA) in a 1" round mold. The resin was cured for 15 minutes with UV light (Jaxman U1c Flood flashlight) at a wavelength of 365 nm. The cross section was ground with 320, 400, 600, and 1000 grit SiC paper (Buehler, Lake Bluff, IL) and polished with 300 nm and then 50 nm alumina suspensions (Buehler, Lake Bluff, IL), dialyzed against 22 g/L $Na_2CO_3$ in DI $H_2O$ solution before polishing, and the 22 g/L $Na_2CO_3$ solution was also added onto the polishing felt regularly during polishing to prevent dissolution of the $CaCO_3$. The polished sample was cleaned using TexWipe® Cotton (Texwipe, Kernersville, NC), air dried, and coated in 20 nm Pt using a Cressington 208 Series sputter coater (Cressington Scientific Instruments, Watford, England) equipped with a thickness monitor while spinning and tilting. The SEM cross-sections were analyzed in backscattered electron (BSE) mode with a Hitachi S3400 variable pressure scanning electron microscope, located in the Ray and Mary Wilcox SEM lab in the University of Wisconsin-Madison Department of Geosciences.

**Table S1:** Resulting SEM MTT and σ measurements compared to fitted values from hyperspectral mapping.

| Site | SEM MTT (nm) | Fitted MTT* (nm) | SEM σ (nm) | Fitted σ (nm) | Number of tablets measured in cross section |
|---|---|---|---|---|---|
| 1 | 391 | 395 | 40 | 40 | 159 |
| 2 | 376 | 390 | 39 | 40 | 182 |
| 3 | 378 | 395 | 38 | 40 | 97 |

* 25 nm is added to all fitted values to account for the organic binding polymer. SEM measurement do not allow us to discriminate between the aragonite and the organic layers.

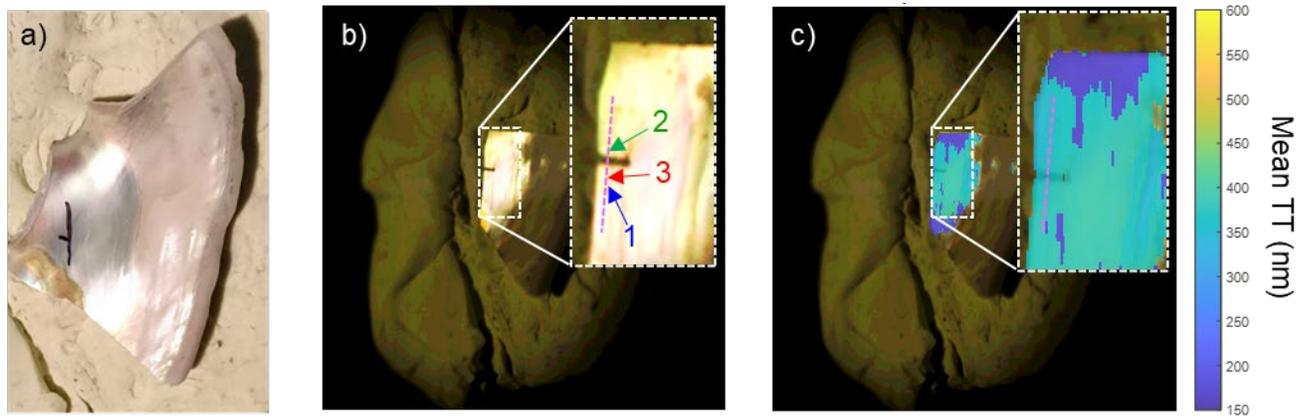

**Figure S9: Cross-sectional scanning electron microscopy validation of nacre MTT and σ.** a) A color image of the sectioned piece of red abalone nacre. The original shell size is 135 mm corresponding to the oldest mollusk in our data set. The black pen mark with notch demarks the region of interest for dicing and sectioning. b) A color-rendered image taken from the hyperspectral map of the nacre post dicing. An approximate 1 mm notch is made to easily identify the measurement location under the SEM. The measurement regions of interest are noted as site 1, 2, 3 and are recessed from the saw edge. The pink dashed line demarks the new edge and region of measurement after embedding the material and polishing the saw mark. c) An overlay of the MTT map on top of the color rendered image. The top 100 – 200 layers from the nacre surface were used to calculate the MTT thickness from the SEM measurements. This is representative of the depth of nacre our hyperspectral mapping method probes.

## 8. Dependence of nacre tablet disorder with age

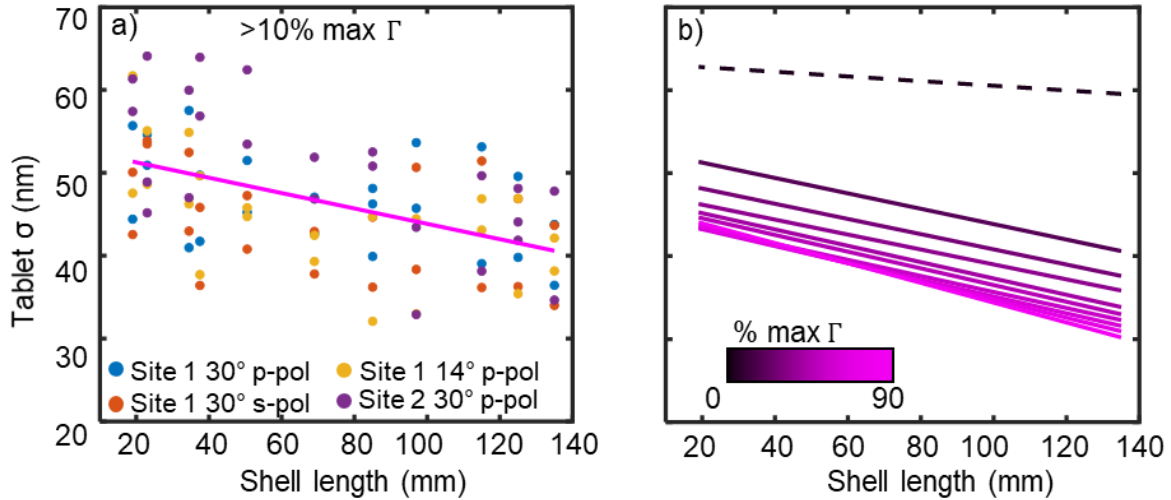

**Figure S10: Ontogenetic dependence of tablet degree of disorder (σ) of red abalone.** a) A plot showing fitted degree of disorder for all sample sites measured across 22 red abalone shells as described in the main article. An inverse relationship with shell size, which correlates to the age of the mollusk, can be seen with the linear fit to the average thickness for each group (pink dashed line). The threshold for data included in this specific plot is for any results with a Γ greater than 10% of the maximum Γ. b) Plots of linear-fit trends obtained selecting different scrubbing thresholds, with black representing unscrubbed data, and the lightest magenta representing datasets scrubbed with a minimum Γ threshold set to 90% of the peak Γ for each map.

## 9. Size and MTT dependence on age

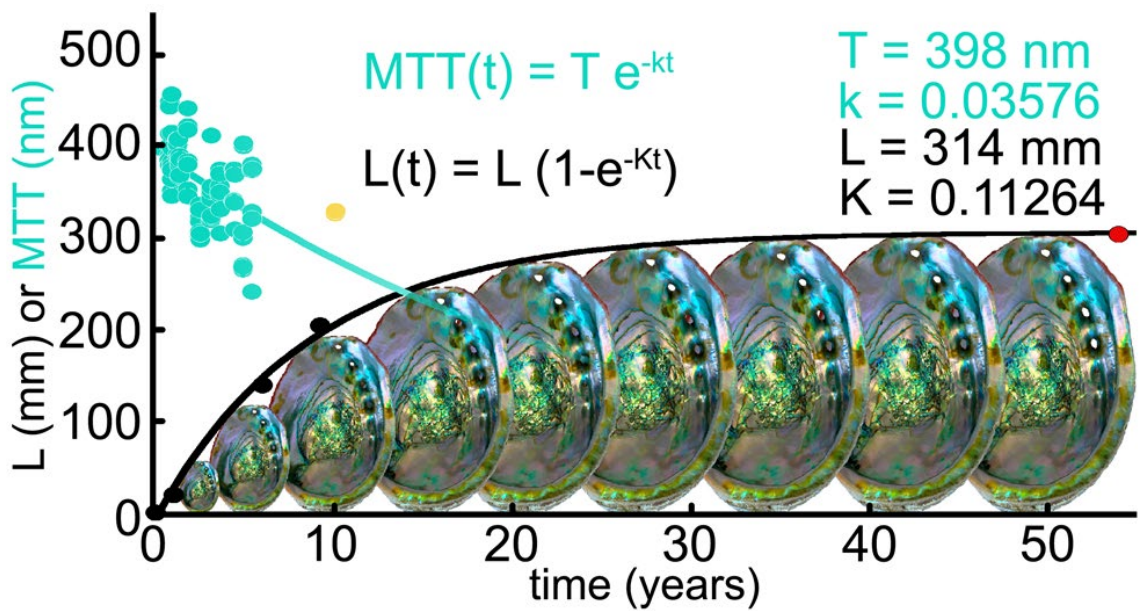

Figure S11: **Red Abalone growth and mean tablet thickness (MTT) dependence on shell age.** A plot shows the shell size dependence on time—or shell age—after fitting to a von Bertalanffy function (black line), which is an exponential rise describing the growth rate of animals. The red dot corresponds to the oldest and largest sized red abalone. In aqua green is the fit of MTT data versus age using an exponential decay. Confidence in the MTT extrapolation decreases as age increases, thus we do not extrapolate far beyond the measured data. The yellow MTT data point is measured from a 205 mm red abalone shell with an unknown growth history and environmental temperature. It was not harvested at the same time or location as all other samples in this work, but is consistent with the other scattered data points and included in the exponential decay fit. A photo of red abalone is repeated and scaled to visually represent the shell size variation with age.

All animals grow faster when they are young than when they are old. This growth behavior follows a von Bertalanffy curve[1], described by the exponential function $L(t) = L[1 - e^{-Kt}]$, where $L(t)$ the length of the shell (or any other animal size parameter), t is time or age, and L and K are species-specific constants. In Fig. S11, the black curve is fit to the von Bertalanffy function using four age datapoints (0, 1, 6, and 54 years). The oldest—thus largest— of our samples was 6 years old. We included a further data point at 54 years based on the oldest age the species lives[2] and the largest size it grows[3] (red dot, Fig. S11). The constants of the fit were L = 314 mm and K = 0.11264.

Using the von Bertalanffy fit, we inferred the ages of all shells measured in this work and plotted their measured MTT versus time (cyan dots, Fig. S11). We fit an exponential function of the form $MTT(t) = Te^{-kt}$ (cyan curve, Fig. S11), where MTT(t) is the mean tablet thickness, t is age, and T and k are fit parameters with values of 398 nm and 0.03576, respectively. The expectations were that red abalone deposited tablet layers of identical thicknesses, and deposited more layers per unit time when they were young and progressively fewer as it grew in size and age. The observation that the thickness of the layers, instead, decreases with age is unexpected and interesting.